\documentclass{article}
\usepackage{ijcai21}

\usepackage{times}
\usepackage{soul}
\usepackage{url}
\usepackage[hidelinks]{hyperref}
\usepackage[utf8]{inputenc}
\usepackage[small]{caption}
\usepackage{graphicx}
\usepackage{amsmath}
\usepackage{amsthm}
\usepackage{booktabs}
\usepackage{amssymb}%打印空心字体
\usepackage{graphicx}
\usepackage{float}
\usepackage{subfigure}
\usepackage[ruled,vlined,linesnumbered]{algorithm2e}
\usepackage[all]{xy}
\usepackage{setspace}

\title{\textsc{AnomalyMaxQ}: Anomaly-Structured Maximization to Query in Attributed Networks}

\author{
Xinyue Zhang$^{1,2}$\and
Nannan Wu$^{\ast,1,2}$\and
Zixu Zhen$^2$\And
Wenjun Wang$^{1,2}$
\affiliations
$^1$State Key Laboratory of
Communication Content Cognition,
China\\
$^2$College of Intelligence and Computing, Tianjin University, China\\
\emails
zhang\_xinyue@tju.edu.cn,
wunannan@act.buaa.edu.cn,
zxzhen@tju.edu.cn,
wwj@pku.org.cn
}

\begin{document}

\maketitle

\begin{abstract}
The detection of anomaly subgraphs naturally appears in various real-life tasks, yet label noise seriously interferes with the result. As a motivation for our work, we focus on inaccurate supervision and use prior knowledge to reduce effects of noise, like query graphs. Anomalies in attributed networks exhibit \textit{structured-properties}, e.g.,   anomaly in money laundering with ``\textit{ring structure}'' property.
%Graphs are useful data model that can naturally represent various entities and their relationships through forming a network. With the wide application of graph database and the growth of the number of nodes, mining potential information in graph data has become an important issue. 
It is the main challenge to fast and approximate  query anomaly in attributed networks. We propose a novel search method: 1) decomposing a query graph into stars; 2) sorting attributed vertices; and 3) assembling anomaly stars under the root vertex sequence into near query. We present \textsc{AnomalyMaxQ} and perform on 68,411 company network (Tianyancha dataset),  7.72m patent networks (Company patents) and so on. 
Extensive experiments show that our method has high robustness and fast response time. When running the patent dataset, the average running time to query the graph once is about 252 seconds.
\makeatletter{\renewcommand*{\@makefnmark}{}
\footnotetext{* The corresponding author is Nannan Wu.}\makeatother}

% In this paper, we address an advanced method to  for optimizing a generic nonlinear cost function subject to a specific query structural constraint, which is defined as follows: given a query graph $\mathcal{Q}$, we locate the closest matchings of $\mathcal{Q}$ in a large data graph $\mathcal{G}$. 
% At the heart of the proposed algorithm are two techniques, including (1) a novel matching strategy of query graph decomposition and candidate extended structure, and (2) linear time scan statistics to find more potential candidate vertices at an early stage.
% The proposed \textsc{AnomalyMaxQ} bears some distinctive features, including (1) \textit{generality}, being able to handle different types of inexact matching (e.g., missing nodes, missing edges, intermediate vertices) on node attributed and/or edge attributed graphs; (2) \textit{efficiency}, our extensive experiment results show that, our approach is more efficiency to existing methods in several real-world anomaly detection tasks.
\end{abstract}

\section{Introduction}
In recent years, high-impact applications, e.g., business investing activities, patent co-author relationships \cite{2020SubgraphMatching}, cheminformatics \cite{EfficientSubgraphMatching}, and complex network  \cite{Willett1998Chemical,Yang2007Path}, are naturally represented as attributed networks \cite{2017efficientAndScalable}. Anomaly detection in attributed networks has attracted much more attention among users in research and industry fields. In supervised learning scenarios, deep learning typically requires a vast number of training data with accurate labels to obtain good performance \cite{deeplearning}. Nonetheless, for a company, such data is barely acquirable due to  artificial compilation. Users usually have the prior knowledge to anomalies, e.g., anomaly in money laundering exhibiting with ``\textit{ring structure}''. Given the ring-query graph, users aim to identify the most anomaly subgraph with the ring structure in attributed networks. In this paper, our target is to present a method that can identify anomaly subgraph isomorphism to the query graph approximately in weak supervision setting using prior knowledge. 
% In this paper, Our target is to present a system that can be fast to identify anomaly subgraph isomorphism to the query graph approximately.

\textit{
\textbf{Example of investment risk anomaly query.} We consider one company as one node and the investment relationship (i.e., the company $ca$ invested in the company $cb$)  as one edge $(ca, cb)$. Each node has the attributes $\mathbf{W}_{ca} \in \mathbb{R}$ (e.g., investment volume, number of accusations or charges,  number of contract disputes). We first build the business attributed networks ${G=(V,E,\mathbf{W})}$ where $V$, $E$, $\mathbf{W}$ represent the node set, edge set, and the attribute set. 
}

\textit{
Then the user has the prior knowledge to ring anomaly-structured investment relationship. We assume the three unknown companies $V=\{ca,cb,cc\}$ have the investment relationship $E=\{(ca,cb),(cb,cc),(cc,ca)\}$. We take $Q=(V,E)$, and  denote $C(ca)=\mathbf{W}_{ca}^t$ as investment volume of the company $ca$ at the time $t$, and $B(ca)=1/|T|\sum_{t\in T} \mathbf{W}_{ca}^t$ as the average investment volume of $ca$ over a period of time $T$. Given the subgraph $S$ and the function $F(S)$, we have $F(S)=C(S) \log{C(S)/B(S)} +B(S)-C(s)$ if $C(S)>B(S)$, and $F(S)=0$ otherwise. These investment structures may be money laundering groups. We are target for the investment risk anomaly in the problem~(\ref{eq:problem-intro})
}
\begin{equation}
    \max_{S \subseteq G} F({S}) \qquad s.t. \quad S \sim Q
    \label{eq:problem-intro}
\end{equation}
where $S \subseteq G$ represents $S$ is a subgraph of $G$, and $S \sim Q$ represents $S$ is approximate isomorphic to $Q$. We can observe that the optimization to $S \subseteq G$ and $S \sim Q$ is NP-hard problem. In this paper, we employ two approaches to relax the hard problem: (1) $S \subseteq G$ is relaxed to $S \subseteq [V]$ select the top $k$ vertices from the sorted $[V]$ as the \textit{upper bound of anomaly}; and (2) $S \sim Q$ is relaxed to $star(S)\sim star(Q)$ where $star$ decomposes $S$ and $Q$ as a sequence of ``star'' like graphs. We assemble  stars into $S$ as the \textit{lower bound of anomaly}. 

For the two approaches, we iteratively update the upper bound and lower bound of the anomaly to achieve the most anomaly-structured maximization to the query graph. We develop the algorithm \textsc{AnomalyMaxQ}, which can be applied to bioinformatics and cheminformatics fields.
%can model compounds and proteins, and graph queries can be used for screening, drug design, motif discovery in protein structures, etc \cite{Efficient_algorithms}. This paper shows a new software system called \textbf{DEMO}. The system combines our algorithm \textsc{AnomalyMaxQ}, Which realizes the function of visible analysis, risk prediction and decision support for datasets.
% In this paper, we propose a novel anomaly subgraph matching algorithm in single-graph database, called \textsc{AnomalyMaxQ}.  Before the formal definition in Section 2, let us see some motivation examples as follows.
% In order to recommend more accurate personnel, we formulate the following specific-shape constrained minimization problem to uncover the specific shape attack subgraph anomalies:
% \begin{equation}
% 	\max \textit{F}(\mathcal{S}) \qquad s.t. \quad \mathcal{S} \in \mathcal{M}
% \end{equation}
% Given the graph $\mathcal{G} = (V,E,\mathbf{W})$ where $\mathcal{S}$ is a subgraph of $\mathcal{G}$, and $\mathcal{M}$ is the family set of   `specific-shape' vertex sets. The symbols in the formula will be explained in a minute. Finally, by optimizing the cost function $F$, we find the matching nodes in Figure 2.

\begin{figure*}[t]
\setlength{\belowcaptionskip}{-0.4cm}
 \centering
 \includegraphics[width=0.9\linewidth]{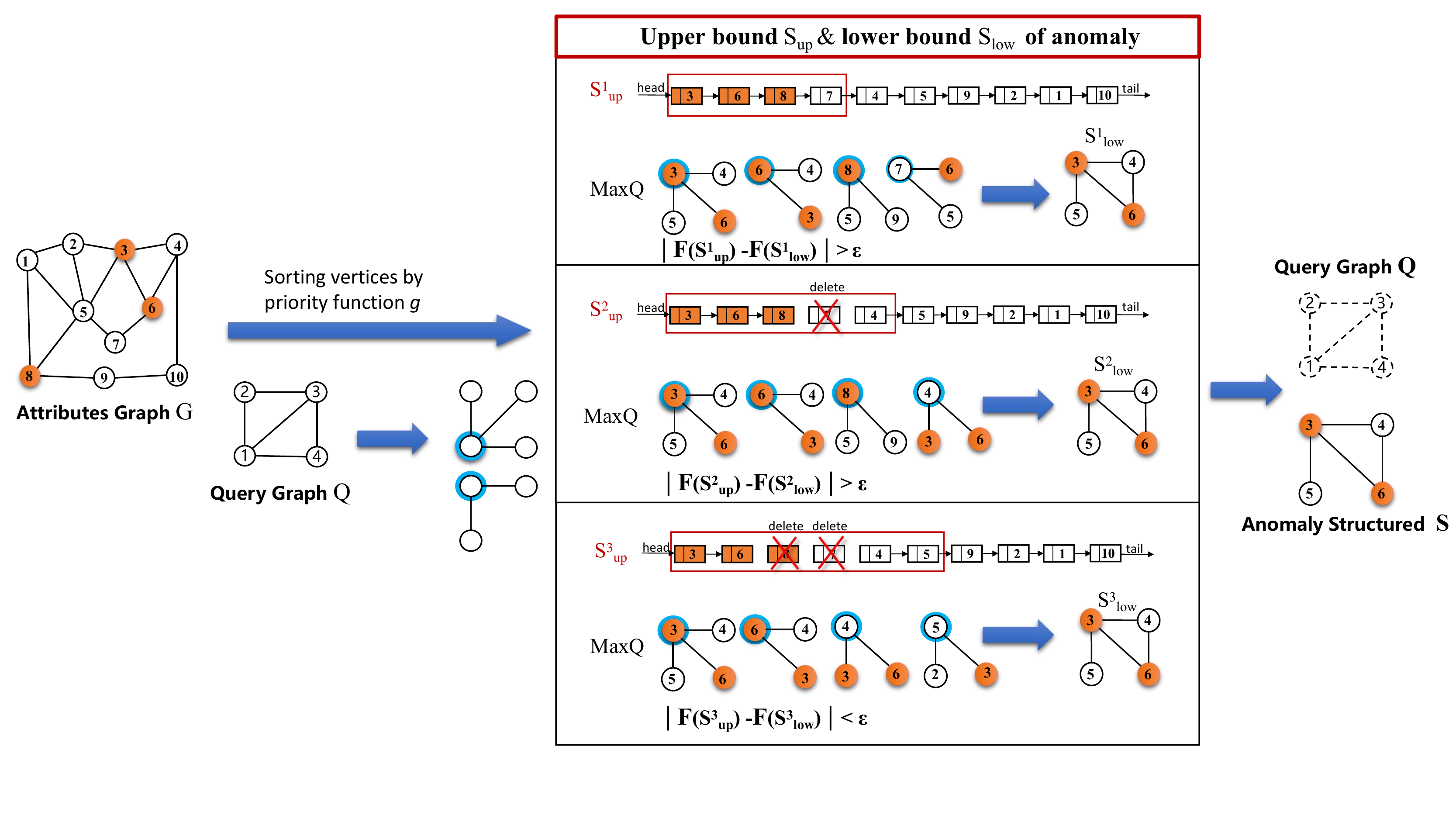}
 \vspace{-9mm}
 \caption{An illustration of our work. This graph shows the process of querying on attributes graph $\mathcal{G}$, similar to query graph $\mathcal{Q}$. The orange node in $\mathcal{G}$ represents the abnormal node, and the white node represents the normal node. And we set $\epsilon$ in our experiments (e.g., $\epsilon = 10^{-6}$)} 
 \label{fig:result-data1}
\end{figure*}

\textbf{Related work.}
% The work is closely related to two aspects: anomaly detection and subgraph matching. 
Anomaly detection assumes that outliers and normal nodes are generated from different distributions.
There are two kinds of anomaly detection methods, including parameterized scanning statistics and non-parametric graph scan (NPGS) statistics~\cite{Anomaly_Detection,2018Futureframe,DBLP:journals/tkde/WuCLHZLR19,DBLP:conf/cikm/00050WYC20}.
% Traditional parameterized scanning statistics need to assume a specific distribution in advance. 
Our work on NPGS, which no longer required assumption of specific forms of node distribution. 
% It compares the current features of each vertex with the historical data of the vertex, so as to estimate the empirical p-value of each vertex, which determines the degree of the anomaly.
Existing subgraph matching work can be grouped exact matching methods \cite{Graph_Indexing,Closure_Tree,GString} and inexact matching methods \cite{G_Finder}.
% We know that graph matching is a NP-hard problem \cite{np_hard}. In fact, it is an interesting challenge that motivate people to design practical algorithms, such as \cite{Graph_Indexing,Closure_Tree,GString}. 
% Consider somtimes do not need exact matching results, 
we propose an algorithm mainly focus on approximate subgraph matching on the large data graph \cite{2019CICE,2019DistributedSubgraph}.

 %\subsection{Contribution}
 %There are several common measures to calculate the similarity of graphs, including graph kernel \cite{graph_kernel,graph_kernal_2}, embedding based graph similarity \cite{graph_embedding} and graph edit distance \cite{ged}. We used to use the modules in networkx \cite{networkx} for calculation. However, with the increase of node size, these modules are not suitable for fast anomaly detection and early warning system. Here we improve the module in Gmatch4py \cite{gmatch4py}. Of course, we will update the system with more efficient algorithms.  
The main contributions of this paper are summarized as follows:
\begin{itemize}
    \item \textbf{Linear-time algorithm}. A novel algorithm is proposed to fast identify anomaly subgraph isomorphism to the query approximately.
    \item  \textbf{Performance}. Extensive experiments on several benchmark datasets demonstrate that the algorithm performs better than the representative methods for this task on both accuracy and run time.
    \item \textbf{Scalability}. Our proposed algorithm is suitable for optimization of a variety of graph scan statistics, which satisfy liner time subset scanning property.
\end{itemize}
%A new algorithm is integrated to support the analysis and prediction system. Over the structure specific model $\mathcal{Q}$, our proposed algorithm is required to minimize loss function $F(x)$ by query graph decompose and linear time subset scanning. Extensive experiments on a number of benchmark datasets demonstrate that the algorithm performs better than the representative methods for this task on both accuracy and run time.
    
 %It has the assets of expansibility and high availability. Our proposed algorithm is suitable for optimization of a variety of graph scan statistics as the target for anomaly specific-shaped subgraph detection.
    
 %Users have a high degree of freedom, and can set the required query graph according to the system recommendation or their own intention.

\section{Methodology}

\subsection{Problem Formulation}
\begin{table}[]
\centering
\small
\begin{tabular}{cp{5cm}}
\toprule
Symbols & Description \\
\midrule
$\mathcal{G}$   & an attributed network         \\ 
$\mathcal{Q}$   & a specific shape query graph     \\ 
$\mathcal{S}$   & a subgraph of $\mathcal{G}$                  \\ 
% $V$   & the node set of $\mathcal{G}$    \\ 
% $E$   & the edge set of $\mathcal{G}$     \\ 
% $S$   & the node set of $\mathcal{S}$     \\
$\mathbf{w}$   & the node attributes         \\ 
$\mathcal{M}$   & the graph-structure model        \\ 
$F$   & the differentiable score funtion \\
$ged(\mathcal{S},\mathcal{Q})$ & the minimum graph edit distance between $\mathcal{S}$ and $\mathcal{Q}$\\
% $P$   & a sequence of edit path          \\ 
\bottomrule
\end{tabular}
\vspace{-1mm}
\caption{Representative symbols} 
\label{tab:booktabs}
\vspace{-5mm}
\end{table}
\textbf{Notation.} First, we briefly review the terminologies that we will use in this paper. Table 1 lists commonly used symbols in this paper. We use italic uppercase letters to denote sets (e.g., $V, E$). The bold uppercase letters are matrices (e.g., $\mathbf{W}, \mathbf{X}$), and the bold lowercase letters are vectors (e.g., $\mathbf{w}, \mathbf{x}$). In this paper, calligraphic letters are networks \footnote{network and graph can be used interchangeably.}.
An \textit{attributed network} $\mathcal{G} = \left (V, E, \mathbf{W} \right )$ consists of: (1) the vertex set $V=[n]=\{1,2,\ldots,n\}$; (2) the edge set $E \subseteq V\times V$, where $|E|=p$; (3) the vertex attributes $\mathbf{W} \in \mathbb{R}^{n \times T}$, where the row vector $\mathbf{w}_{v} \in \mathbb{R}^{T}$ is the attribute values observed within the time span $T$ for the vertex $v \in V$. For the node subset $S \subseteq V$, $\mathbf{W}_{S} \in \mathbb{R}^{|S| \times T}$ keeps row vectors only in the set $S$.
% For a graph $\mathcal{S}$, we sometimes write $V_{\mathcal{S}}$, $E_{\mathcal{S}}$ and $\mathbf{W}_{\mathcal{S}}$ if its vertex set, edge set and attributes are clear from context.
We define subgraph $\mathcal{S}$ of $\mathcal{G}$ as $\mathcal{S} \subseteq \mathcal{G}$ if $V_{\mathcal{S}} \subseteq V, E_{\mathcal{S}} \subseteq E$ and  $\mathbf{W}_{\mathcal{S}}$ is restricted from $\mathbf{W}$. Let $\mathcal{Q}$ be a query graph. 

%\textbf{Definition 1.  \: Linear Time Subset Scanning (LTSS)}

%LTSS is a powerful and useful tool which allows extremely effificient unconstrained optimization over all subsets. \shortcite{LTSS} formally defined the LTSS property and demonstrate that a large class of functions and several commonly used spatial scan statistics satisfy this property. Let $\mathcal{G} =\left \{R_{1},\dots, R_{n}\right \} $ be a set of $n$ data records, and let $F\left ( \mathcal{S}\right )$ be a cost function mapping a subset of data records $\mathcal{S}\subseteq \mathcal{G}$ to a real number. 
%We refer to $g$ as a \textit{`priority function'}, and $g\left ( R_{i}\right )$ as the \textit{`priority'} of data record $R_{i}$. Next we define $R_{\left ( k\right)}$, $k = \left \{1,\dots,n\right \}$, to be the data record $R_{i}\in \mathcal{G}$ with the $k$-th highest value of $g\left ( R_{i}\right )$ means the `$k$-th highest priority record'. For a given data set $\mathcal{G}$, the score funtion $F\left(\mathcal{S}\right)$ priority function $g\left ( R_{i}\right)$ satisfy the LTSS property if and only if:
We refer to $g$ as a \textit{``priority function''}. The score function $F\left(\mathcal{S}\right)$ and priority function $g$ satisfy the Linear Time Subset Scanning (LTSS) property~\cite{LTSS} if and only if:
\begin{equation}
\max_{S\subseteq \mathcal{G}}\left \{F\left ( \mathcal{S}\right )\right \}=\max_{k=\left \{1,\dots, n\right \}}\left [ F\left ( \left \{V_{\left ( 1\right )}\dots V_{\left ( k\right )}\right \}\right )\right ]
\end{equation}
If $V$ is already sorted by priority with its record data, this property allows us to maximize $F\left ( \mathcal{S}\right )$ in $O\left (N\right )$ time. Otherwise, we must first sort the records by priority, which requires $O\left ( N\log\left ( N\right )\right )$ time.

\textbf{Problem: Anomaly-Structured Maximization (ASM)}
Given an attributed network $\mathcal{G}$, and the anomaly-structured query graph $\mathcal{Q}$, we maximize the score function over all subgraphs in attributed networks:
\begin{equation}
    \max_{\mathcal{S} \subseteq \mathcal{G}} F(\mathcal{S}) \qquad s.t. \quad \mathcal{S} \sim \mathcal{Q}
    \label{eq:problem}
\end{equation}
where $F(\mathcal{S})$ can employ score functions that satisfy the LTSS property, e.g., \textit{Kulldorff's\;(KULL)} original spatial scan statistic, \textit{Expectation-based Poisson\;(EBP)} scan statistic.

% \begin{theorem}[\textbf{Graph-Structured Model (GSM)}]
% \end{theorem}
% Given an attributed network  $\mathcal{G}=(V,E,\mathbf{W})$ and the anomaly-structured query graph $\mathcal{Q}=(V_{\mathcal{Q}}, E_{\mathcal{Q}})$  where $|V_{\mathcal{Q}}|=m$, then the graph-structured model $\mathcal{M}(\mathcal{Q})$ can be defined in
% \begin{equation}
%     \mathcal{M}(\mathcal{Q}) = \{\mathcal{S} : \mathcal{S} \subseteq \mathcal{G}, ged(\mathcal{S}, \mathcal{Q}) \leq e \}
%     \label{eq:model}
% \end{equation}
% where $e$ is the budget on error tollerance of graph edit distance ($ged$) between $\mathcal{S}$ and $\mathcal{Q}$. The four operations of $\mathcal{S}$  transforming to the query graph $\mathcal{Q}$ consist of inserting and deleting one node and edge, whose each cost of the four operations is one. An graph edit path is a sequence of edit operations $P=<p_1,p_2,\ldots,p_k>$ from $\mathcal{S}$ transforming to $\mathcal{Q}$ (e.g., $\mathcal{S} = \xymatrix@1{\mathcal{S}^0\ar[r]^{p_1}_{} &\mathcal{S}^1  \ar[r]^{p_2} & \ldots \ar[r]^{p_k}& \mathcal{S}^k \cong \mathcal{Q}}$). The graph edit distance is $ged(\mathcal{S},\mathcal{Q})=\sum_{i=1}^k c(p_i)$ for the cost $c(p_i)=1$. The subgraph $\mathcal{S}$ is isomorphic to $\mathcal{Q}$ if $e=0$. 

\subsection{AnomalyMaxQ}
The overall idea of this method is to iterate the subgraphs of upper and lower bounds in Algorithm 1. 
\begin{algorithm}[t]
\small
% \begin{spacing}{1.5}
\label{alg:algorithm}
\SetAlgoLined
\KwIn{Query $\mathcal{Q}$, attributed network $\mathcal{G}=(V,E,\mathbf{W})$}
\KwResult{Anomaly-Structured $S_{low}$ }
 
 $ (v_{(1)},v_{(2)},\ldots,v_{(n)}) \leftarrow g(\mathcal{G}), i \leftarrow 0, S_{up}^i, S_{low}^i \leftarrow  \{\} $\;

 \Repeat{$\left | F\left (S^i_{up}\right )-F\left (S^i_{low}\right )\right |< \varepsilon $}{
 
 $v_{(j)} \leftarrow \max g\left ( S_{up}^{i} \cap S_{low}^{i}\right )$ \; 
 
 $v_{(k)} \leftarrow \min g(S^i_{up})$ \;
 
 $S \leftarrow S_{up}^{i} \cap S_{low}^{i} $ \;
 
 $S_{up}^{i+1} \leftarrow S \cup \max g \big(\{v_{(j)}, \ldots, v_{(k+1)}\} - S, m-|S|\big)$ \;
 
 $S_{low}^{i+1} \leftarrow \mathrm{MaxQ}(S_{up}^{i+1}, \mathcal{Q})$ \;
 
 $i \leftarrow i + 1$ \;
 }
 \textbf{return}{\: $S_{low}^{i}$;}
 \caption{\textsc{AnomalyMaxQ}}
%  \end{spacing}
\end{algorithm}
First (Root Selection, Line 1), \textsc{AnomalyMaxQ} evaluates the matching priority of vertices based on their empirical p-values. Second (Upper-Bound Structure Construction, Lines 3-6), a dynamic filtering and refinement strategy is used to maximizing score function of attribute graph $\mathcal{G}$. Finally (Lower-Bound Structure Construction, Line 7), \textsc{AnomalyMaxQ} searches Lower Bound Graph Structure  according to the graph edit distance from the query graph to return the approximate optimal results which have the least loss function cost. In the remaining of this section, we highlight each of these three steps.

% First (Root Selection, Line 1), \textsc{AnomalyMaxQ} construct the matching order of the vertices based on their empirical p-values. Second (Upper-Bound Structure Construction, Lines 3-6), a dynamic filtering and refinement strategy is used to maximizing score function of $\mathcal{G}$. Finally (Lower-Bound Structure Construction, Line 7), \textsc{AnomalyMaxQ} searches Lower Bound Graph Structure  according to the graph edit distance from the query graph to return the approximate optimal results which have the smallest loss function cost. 
% In the remaining of this section, we highlight each of these three steps.

%\subsubsection{Root Selection}
%Given a attributes graph $\mathcal{G}$, we first select m root vertices to start the matching process, where $m$ is the number of nodes in the query graph $\mathcal{Q}$. Basically, we would like to choose them which (1) have as few candidates as possible, and (2) are the most anomalistic vertex. With these two design objectives in mind, we first construct the matching order of the vertices according to priority function.  For example, in Figure 1, we choose $\left\{v_{3},v_{6},v_{8},v_{7}\right \}$ as the root. 
% In Section 2, we prove that several commonly used spatial scan statistics satisfy the LTSS property with priority function equal to the empirical p-value. The empirical p- value is calculated as follows: if the historical data feature of this vertex is larger than the current feature, then the outlier is added by one; the accumulated outlier is divided by the total number of days to get it.

\textbf{Root selection.}
Given an attribute graph $\mathcal{G}$, we first select $m$ root vertices to start the matching process, where $m$ is the number of nodes in query graph $\mathcal{Q}$. We would choose nodes which (1) have as few candidates as possible, and (2) have most anomaly vertices. 
% With these two design objectives in mind, we first construct the matching order of the vertices according to priority function.  
% In Section 2, we prove that several commonly used spatial scan statistics satisfy the LTSS property with priority function equal to the empirical p-value. The empirical p- value is calculated as follows: if the historical data feature of this vertex is larger than the current feature, then the outlier is added by one; the accumulated outlier is divided by the total number of days to get it.
For example, in Figure 1, we can choose $\left\{v_{3},v_{6},v_{8},v_{7}\right \}$ as the root. \textbf{Upper-bound of anomaly}.
%Following the root selection, the next step is to construct the upper-bound graph structure $S_{up}$ by updating the vertices. Note that $\mathcal{S}_{up}$ may not need to be connected, even if they are isolated points.
%The key of this step is the update strategy of the node, deciding the selection of the candidate upper-bound vertices. We want to make sure that $\mathcal{S}_{up}$ is composed of $m$ nodes each time. 
It consists of two parts: 1) reserve the previous  vertex set $\mathcal{S}$, and 2) add next new vertices. We get the vertex set $\mathcal{S}_{up}$,  which is the upper bound score of anomaly without structure property. We must select vertex not computed to ensure $\mathcal{S}_{up}$ will not get stuck in an infinite loop.
%In particular, when the number of vertex in $\mathcal{S}$ is equal to $m$, $\mathcal{S}$ is returned as result, or  we update $\mathcal{S}$ in descending order. As shown in Algorithm 1 (line 6), assume that the iteration reaches the $k$-th vertex $v_{(k)}$ and $\left | S\right|$ vertices are retained. To ensure $\mathcal{S}_{up}$ will not get stuck in an infinite loop, we must slect $v_{(k+1)}$, then add $m-\left | S\right |-1$  nodes to update loop. For example, in Figure 2, $\left\{v_{(3)},v_{(6)},v_{(8)},v_{(4)}\right \}$ will be the upper bound structure of the next iteration.
% We will select $m-\left | S\right |-1$ vertices in order after $v_{(j)}$, and we must select $v_{(k+1)}$ to ensure $\mathcal{S}_{up}$ will not get stuck in an infinite loop. 
% $\left \{v_{(3)},v_{(6)}\right \}$ is the vertex set needs to be preserved.
% Then, $\left\{v_{(3)},v_{(6)},v_{(8)},v_{(4)}\right \}$ will be the upper bound structure of the next iteration.
\textbf{Lower-bound of anomaly}.
%In this step, we construct the lower bound structure $\mathcal{S}_{low}$ based on the node set of the upper bound structure $\mathcal{S}_{up}$, the details of which  as can be seen in Algorithm 2. 
We consider $S_{up}$ as root to select star subgraphs, which are isomorphic to stars in the query graph. The matched stars are assembled into one subgraph which are approximate to the query.
\subsection{Theoretical analysis}
\begin{algorithm}[t]
% \begin{spacing}{1.5}
\small
\SetAlgoLined
\KwIn{$\left \{S_{1}^{\ast},S_{2}^{\ast},\dots,S_{m}^{\ast}\right \} \leftarrow Q$,\, $\mathcal{G}$,\, $\mathcal{S}_{up}^{i}$}

\KwResult{lower-bound Anomaly-Structured $\mathcal{S}_{low}^{i}$ }

\For{$v\in S_{up}^{i}$}{
$S_{v}^{k} \leftarrow \underset{S\subseteq Star(v),\:S\cong S_{k}^{\ast}, \: k\in \left \{1,\dots,m\right \} }{\max} F\left (S \right )$;
}
$j\leftarrow 0, \: S^{j}\leftarrow \left \{\right \}$;

\Repeat{ $ged\left ( S^{j},Q\right )> ged\left ( S^{j-1},Q\right )$ }{
$S^{j+1}\leftarrow  \underset{v\in S_{up}^{i}, \: k\in \left \{1,\dots,m\right \}}{\arg\min}ged\left ( S^{j} \cup S_{v}^{k},Q\right )$;
$j \leftarrow j+1$;
}
\textbf{return}{\: $S_{low}^{i}\leftarrow S^{j-1}$;} 
\caption{\textsc{MaxQ}}
% \end{spacing}
\end{algorithm}
\textbf{Spatial complexity analysis.} In each iteration for numerical conjunctive stars~\cite{DBLP:journals/tkde/ZhaoCY20}, we only keep $\mathcal{S}_{up}$ and $\mathcal{S}_{low}$ in step 6 of Algorithm 1. The space complexity of \textsc{AnomalyMaxQ} is $O\left (m\right )$.  The size of two subsets is determined by the number of nodes $m$ of query graph $\mathcal{Q}$.

\textbf{Time complexity analysis.} The time complexity of the algorithm is mainly determined by \textsc{MaxQ} in step 7. \textsc{MaxQ} matches the subgraph of each node and its neighbors in $\mathcal{S}_{up}$  with the decomposed query graph to calculate the largest isomorphic part. According to the worst case, the decomposed query graph can have $m$ kinds of neighbors, so each node in $\mathcal{S}_{up}$  has $m$ matching methods at most. In general, the time complexity of the algorithm is $O\left( N \times M\right )$.

\section{Experiments on Real Datasets}
\subsection{Experiment Design}
To verify the performance of \textsc{AnomalyMaxQ} approach, we conducted experiments on large-scale  artificial toy data and real datasets. 
% We're going to focus on real-world datasets: Tianyancha Dataset and Company Patents Dataset. 
% The following table gives a brief summery of them.

\textbf{Internet Traffic Network.\footnote{The company has more than 0.6 billion users.}} The real-world \textbf{*edu.cn} network dataset consists of \textbf{8,540,966} web sites browsing logs from May 31, 2014 to May 13, 2015. The network  with 31,238 vertices and 118,708 edges was built from the browsing logs (i.e., the edge (IP site A, IP site B) denotes that A visited B). 
% For a day $t$ and a node $v$ in this network, we denote the number of logs within $v$ on that day $t$ as the \textit{observed value} \textbf{c}$_{v} = \textbf{W}^{0}_{v}$, and we calculate the empirical p-value by comparing it with historical data. 
The p-value of vertices are under 0.15 if they were attacked, otherwise were not.  
% We select the subgraph with the highest function score anomaly structure as the approximately optimal result from many subgraphs similar to the query graph. Significantly, we distinguished between the record of normal access and actual attacks. Therefore we  know the attack type, attack time, and IP address location. 
% In other words, we could compute the corresponding precision rate and other calculation indicators according to the calculation results to compare the algorithm. 
For testing the robustness of methods, we flipped p-values of $K \in \left \{5,10,20\right \}$ percent nodes randomly. 
\textbf{Tianyancha Dataset.\footnote{Downloaded from \url{https://www.tianyancha.com/}}}  This graph has 68K nodes, each representing an enterprise in Tianjin. An edge represents the investment relationship between two entities. It contains 54 Tags, such as investing information, legal disputes, etc.
% The data is loaded into the graphics database according to the specific format. We set specific query graph according to our experience. Of course, users can customize it according to their own preferences. In the aspect of calculating empirical p-value, we assume that there are $n$ violation records of company $v$ in a year (such as administrative penalty, legal dispute, tax arrears announcement, abnormal operation, etc.). If the historical value is greater than $n$, then the \textit{observed value} $\textbf{c}_{v}$ plus one, and finally divided by the average time period, the empirical p value can be obtained.
\textbf{Company Patents.\footnote{Downloaded from \url{http://www.wanfangdata.com.cn/}}} There are 7.72m nodes and 1.87m edges among them. The patent dataset includes the patent information in China from 1990 to 2020. We consider each inventor and company as a node. Thus an edge represents connection between the inventor and his company. 
% When a patent has the signatures of two or more companies, an edge is established between the inventor and these companies. The data set contains 7,729,373 points and 1,875,139 edges. 
% When using nonparametric method to calculate anomalies, we use the number of patents published by a company in the latest year as a baseline $\textbf{b}_{v}$ and compare it with its own historical data \cite{Nonparametric}. For a year $t$ and a node $v$ in this network, we denote that if the number of patents on year $t$ less than $\textbf{b}_{v}$, the \textit{observed value} $\textbf{c}_{v}$ plus one. The final the \textit{observed value} divided by the statistical year is the empirical p-value. 
\textbf{Respiratory Emergency Department (ED) Dataset.} We simulate respiratory medical record data sets with different number of nodes, which is $10^{2}$,$10^{3}$,$10^{4}$,$10^{5}$,$10^{6}$ respectively \cite{graphtpp,DBLP:journals/geoinformatica/ZhaoCCJWLR20}. The sparsity setting of the graph is 0.4.
% For each node $v\in V$, we consider each node as a zip code and collect the number of cases in $v$ in 28 days. In the non outbreak period, we assume that the number of cases follows Poisson distribution \cite{2009Neil}. During the outbreak of infectious diseases, we inject Possion $\left ( \left ( T-t\right )w_{v}\bigtriangleup \right )$ cases into infected nodes, where $w_{v}$ denots is the regularization term in infected addresses and $\bigtriangleup$ denotes the outbreak severity.

\textbf{Query-map Setting.} We set six query graphs in Figure 2 for each group of data through the work of others. Other graphs can be made up of these basic shapes or close to them.
% and the number of nodes in the query graph increase gradually.
These  query graphs can be replicated as hundreds of nodes. We use different shape constraint subgraph to try to find out the anomaly form of the data set. Specifically, Figures 2(1) is a ring-shaped network \cite{Top-k_subgraph_matching,Efficient_subgraph_matching}, which represents a state that anomaly subgraphs are interconnected; Figure 2(2) is a linear network graph \cite{querymap}, representing the incident effect point to point, like Water Pollution Cases; Figure 2(3),(6) are star-shaped subgraphs \cite{graphtpp} denote that a node infects its neighbors, then the infection from one star-shaped area to another neighbor. Figure 2(4) is a bipartite graph \cite{graphtpp}, which means a case of many-to-many communication. And Figure 2(5) is a tree-shaped graph \cite{SIIMCO,Turboiso} for detecting anomaly subgraphs have a  superior-subordinate relationship in the group.

\begin{table}[]\small
\centering
\scalebox{0.85}{
\begin{tabular}{lccrr}
\toprule
Dataset                       & \multicolumn{2}{c}{Graph} &  size  &        \\ 
\midrule
Computer Access   &node  & IP Address          & 31,238  &            \\ 
                 &link  & Attack/Access      & 118,708 &            \\ 
Tianyancha   &node  & Company          & 68,411  &            \\ 
                 &link  & Investment relationship      & 151,591 &              \\ 
Company Patents      &node  & Inventor/Company &  7,729,373    &   \\
              & link  & Coauthor/Affiliation         &1,875,139&        \\ 
ED Dataset      &node  &  Address &  1,000,000    &   \\
              & link  &     Contiguous area        &3,996,000 &        \\ 
\bottomrule
\end{tabular}
}
\caption{Summary of two real datasets and a simulated ED dataset} 
\vspace{-1em}
\label{tab:booktabs}
\end{table}
% Tianyancha enterprise dataset and China enterprise patent dataset.

\subsection{Methods}
\textbf{Our Method.} As mentioned in Section 3, our method performs subgraph matching in three steps: 1) query decomposition, 2)  select the candidate nodes and construct upper bound and lower bound structure, and 3) calculate the graph edit distance between the subgraph and the query graph. Query graph employs the nonparametric scan statistic BJ \cite{BJ} and HC \cite{HC} as the objective function to detect specific shape attack subgraphs in the real networks. When the score of upper and lower limits is less than the threshold, the output result finally gets anomaly structured. 

\textbf{Comparative Methods.} Several different techniques have been developed for anomaly detection in many real world scenarios, including 
% NPHGS \cite{NPHGS},
TSPSD \cite{TSPSD}, Graph-TPP \cite{graphtpp}, Query-map \cite{querymap} and so on. The baselines are designed for specific shape anomaly discovery in attributed graphs. Source codes of the baseline methods are provided by the original authors. We followed strategies recommended by them to adjust the related model parameters. The algorithm TSPSD is designed in a nonparametric statistical framework, and the specification of parameters are hence relatively straightforward. We set $\alpha_{max}$ and the number of seed entities $K$ to 0.15 and 5 respectively. TSPSD chooses Steiner Tree heuristics for output. Because of it consider just the connected subgraph anomaly without the specific shape anomaly prior. Therefore, we find the subgraphs that are most similar to the query graph for comparison with others. 

\textbf{Performance metrics.}  We use the following performance metrics: 1) precision. We compute precision of our result, i.e. the ratio of the number of correct anomalous nodes and the number of nodes. The recall metric is ignored for computing as target subgraphs return fixed size of nodes. 2)  running time. The optimization power of our method can be examined in the iteration of graph scan statistics scores. We compare it with baseline methods on running times. 
% Moreover, we denote Graph edit distance (GED) to recognize anomalies patterns with different data sets.

\subsection{Experiment Results}
\begin{figure}[t]
\vspace{-0.8cm}
 \centering
 \includegraphics[width=0.7\columnwidth]{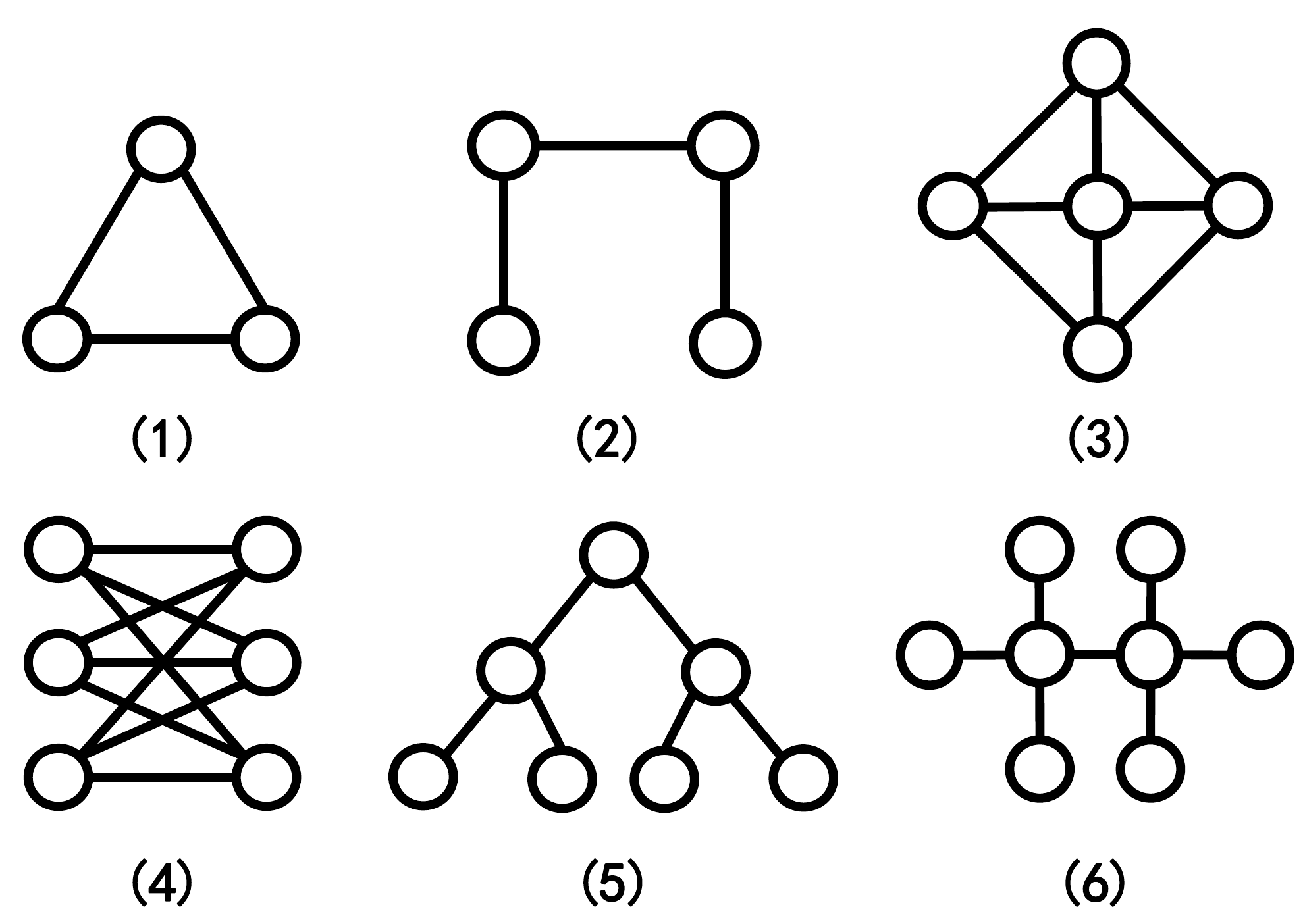}
 \caption{Examples of query graphs we set up. Other graphs can be made up of these basic shapes.}
 \label{fig:result-data1}
 \vspace{-2mm}
\end{figure}

\begin{table*}[]
\vspace{-4.5mm}
\begin{spacing}{1.0}
\centering
\scalebox{0.85}{
\begin{tabular}{ccccccc}
\toprule
Methods   &  $\mathcal{Q}_{1}$    & $\mathcal{Q}_{2}$      & $\mathcal{Q}_{3}$ &  $\mathcal{Q}_{4}$     & $\mathcal{Q}_{5}$               & $\mathcal{Q}_{6}$       \\ 
\midrule
  AnomalyMaxQ(\textbf{BJ})       & (\textbf{1.00}, 0.67, 0.67)                   & (\textbf{1.00}, 0.50, 0.75)              & (\textbf{1.00}, 0.80, 0.70)                 & (\textbf{0.83}, 0.67, 0.67)                                 &(0.71, 0.57, 0.71)                   & (\textbf{1.00}, 0.63, 0.72)                         \\ 
 AnomalyMaxQ(\textbf{HC})   & (1.00, 0.67, 0.67)                   & (1.00, 0.50, 0.50)                             & (1.00, 0.80, 0.70)       & (0.83, 0.67, 0.67)                              &(0.71, 0.57, 0.57)                   & (1.00, 0.63, 0.72)                  \\ 
% NPHGS(\textbf{BJ})     & (1.00, 0.50, 0.67)                  & (1.00, 0.63, 0.75)                                & (0.80, 0.80, 0.80)        &(0.67, 0.75, 0.67)                            &(0.64, 0.79, 0.64)       &  (0.63, 0.63, 0.50)                    \\
% NPHGS(\textbf{HC})    & (1.00, 0.50, 0.67)                  & (1.00, 0.63, 0.75)                             & (0.80, 0.80, 0.80)      &(0.67, 0.75, 0.67)                              &(0.64, 0.79, 0.64)       &  (0.63, 0.63, 0.50)                    \\ 
TSPSD(\textbf{BJ})    & (0.33, 0.33, 0.50)                  & (0.25, 0.25, 0.38)                                & (0.20, 0.20, 0.30)         & (0.17 ,0.17, 0.25)        & (0.14 ,0.14, 0.21)        &  (0.12 ,0.12, 0.19)                      \\
TSPSD(\textbf{HC})    &(0.33, 0.67, 0.33)                   & (0.25, 0.50, 0.25)                                & (0.20, 0.40, 0.20)      & (0.17 ,0.33, 0.17)           &  (0.14 ,0.33, 0.31)                  &   (0.12 ,0.25, 0.25)                        \\ 
Graph-TPP & (0.50, 0.50, 0.55)    &(0.62, 0.62, 0.41)     &(0.20, 0.20, 0.46) & (0.58, 0.58, 0.61)            & (0.57, 0.57, 0.47)               &(0.43, 0.50, 0.58)               \\ 
Query-map & (0.50, 0.50, 0.33) & (0.83, 0.75, 0.83) & (0.27, 0.20, 0.46) & (0.58, 0.58, 0.46) & (0.57, 0.57, 0.47) & (0.43, 0.50, 0.58)          \\ 
\bottomrule
\end{tabular}
}
\caption{Query graphs from $\mathcal{Q}_{1}$ to $\mathcal{Q}_{6}$. Comparison on the precision of structure-specific anomalous subgraphs discovered by methods, run times and graph edit distance, and we put in parentheses the results for \textbf{5\%, 10\%,  and 20\%} noise.}
\end{spacing}
\end{table*}
\vspace{-0.3mm}
\begin{figure*}[h]
\setlength{\belowcaptionskip}{-0.4cm}
 \centering
 \includegraphics[width=0.9\linewidth]{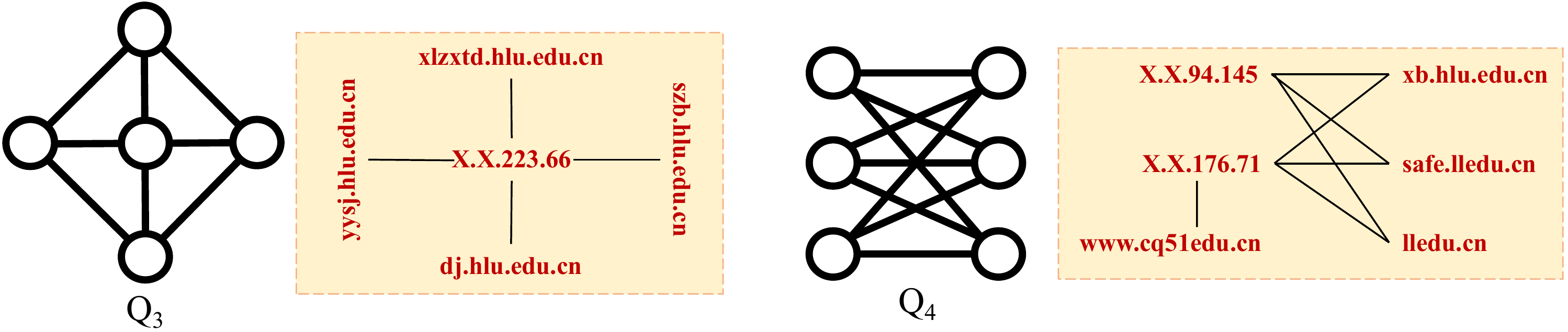}
 
 \vspace{-1mm}
 \caption{The cyber-attacks on March 12, 2015 are detected by our method. Red nodes denote the attacking or attacked IP sites in real-world, and yellow area is the anomaly vertices we calculated.} 
 \label{fig:result-data1}
\end{figure*}

We tested our method \textsc{AnomalyMaxQ} on Internet Traffic Dataset for the star and bipartite-shaped attack patterns. Our algorithm often better able to make connections that were hidden in the jumble of information. Although these IP addresses appear in different places and periods, their attack behaviors are similar. \textsc{AnomalyMaxQ} can obtain some abnormal IP groups by querying specific attack patterns, such as star and bipartite-shaped graph. Our methods successfully discovered the cyber-attack networks without innocent nodes. 
\textbf{Star-shaped attacks.}\; As illustrated in Figure 3, we can see clearly that two attack cases were found by star query graph. On March 10, 2015,  \textsc{AnomalyMaxQ} found the client X.X.223.66 attacked the other server sites \url{yysj.*.edu.cn} and \url{szb.*.edu.cn}; however, the server site \url{www.*.edu.cn} was attacked by four clients on March 12, 2015, different from the last one. These network attack patterns are the most common forms in the network. \textbf{Bipartite-shaped attacks.}\; Figure 3 shows \textsc{AnomalyMaxQ} detected some cyber-attack networks. Because attackers often do not use only a single IP address for cyber-attacks. Compared with the star subgraph, we can discover attack group in the meantime.
% such as the variations of X.X.223.X contributed to many cyber-attacks. 
By recording these IP addresses, we find that these IP addresses come from multiple fixed network segments, and the attack mode and location remain unchanged, which means that these IP addresses may come from the same attack source. With this information, we can prevent attacks by blocking the IP of these fixed IP segments.

\textbf{Precision for target subgraphs detection.} We randomly selected the average accuracy of two days as a result. Table 3 shows the results of subgraph isomorphism search performance using specific shape query graphs for the Internet traffic dataset. We present a comparison of precision for methods under different noise conditions in detail. At 5\% noise level, our proposed \textsc{AnomalyMaxQ} (i.e., $\varphi_{BJ}$ and $\varphi_{HC}$) achieved higher precision (close to 1) than competitive baselines (close to 0.7). Moreover, even at 20\% noise level, it achieved at least 0.70 precision, and baselines achieved the best precision to 0.50.

\subsection{Case Study}

% Although the accuracy of nphgs is also very high, it returns abnormal points without structure, which is very different from the query graph. 

% \textbf{Scalability analysis of running time.} Figure 4 reports the comparison between our methods \textsc{AnomalyMaxQ} and the competitive baseline methods on the running time. In Figure 4, the running times were collected from the computer with \textsc{AnomalyMaxQ}. With the increase of the number of query graph nodes, the running time of subgraph matching is also increasing. The result show that our proposed method ran faster than all the baseline methods in most settings, when the results converge. It is worth mentioning that if we use \textit{python-networkx} to find the isomorphic part of the result and query graph, then the running time is too long to get the result.

% \textbf{Structural similarity.} Downside sub-figures of Figure 4 depicts the graph edit distance between result and query graph. We can observe that star-shaped graph are easier to find. If we have good prior knowledge, the result will have high accuracy and low graph editing distance.

 \textbf{Investment decision support.} 
%  With the increasingly complex ownership structure of modern companies, global companies are entering a period of high incidence of mergers and acquisitions, and capital operation methods and methods are emerging one after another. Especially after the reform of non-tradable shares, the supervision of capital operation in the securities market is becoming more and more difficult.
% Cross shareholding is not only a common way of capital operation, but also a common means of development and expansion. However, the establishment of the relationship between legal persons holding shares would inflated capital or evade tax. Because of the above reasons, it is difficult for VC firms to observe the real operation of the enterprise. 
 Cross shareholding would inflated capital or evade tax. Figure 4 shows the results on Tianyancha dataset, including five enterprises. According to the dataset, we can see that Taikete was established in August 2000 and invested by Tiandi Weiye; Fuwo established in December 2001 and invested by Taikete and Tiandi Weiye; Xinmao was established in July 2016, then other companies invested in it. We also examined the legal risks of these companies. Xinmao is involved in two sales contract disputes, both of which are defendants. Tiandi Weiye involves total of 24 disputes, include housing rental contract disputes, construction project contract disputes, sales contract disputes, labor disputes and so on. Last but not least, as the defendant, Changfei Xinmao was also charged with seven labor disputes.
% \vspace{-2mm}
\begin{figure}[h]
\centering
\setlength{\belowcaptionskip}{-0.4cm}
\includegraphics[width=0.9\linewidth]{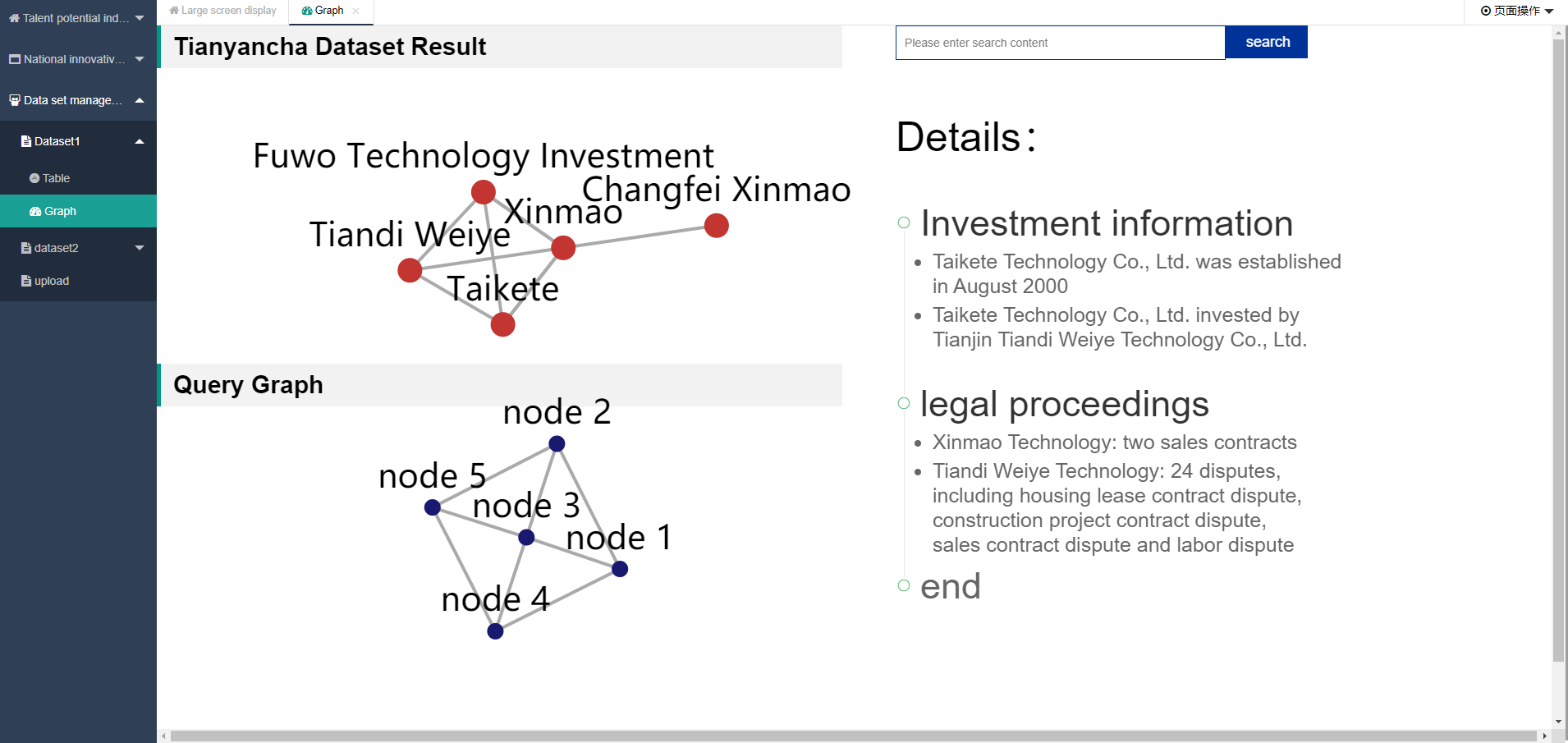}
\caption{Calculation results of patent disputes.}
\end{figure}
% \vspace{-0.3mm}

\textbf{Business Competition Forecast.}
% Around 275,900 PCT international applications were filed in 2020, up 4\% on 2019 despite the global pandemic. When the inventor leaves the original company, the company will sign a competition agreement to avoid the risk of disclosure of trade secrets, but such disputes still occur from time to time. We need an efficient means to detect whether there are companies competing in the same field to prevent the disclosure of their business secrets.
After calculation, we find that Nanjing Melander Medical Technology Company and Nanjing Weisi Medical Technology Company are consistent with our query graph pattern. Melander is prosecuted by Weisi, whose employees hopping to Melander. There is an identified competitive relationship between the two companies. Through China's judicial document website, we find that there is more than one patent dispute lawsuit between two companies. Melander was sued for its patent content (No.: 201320752362.7) was as same as the research of Weisi.
% The patent (No.: 201320752362.7) sued by Weisi on June 12, 2015 which is the same as the research content.

\begin{figure}[h]
\setlength{\abovecaptionskip}{0.1cm}
\centering
\includegraphics[width=0.9\linewidth]{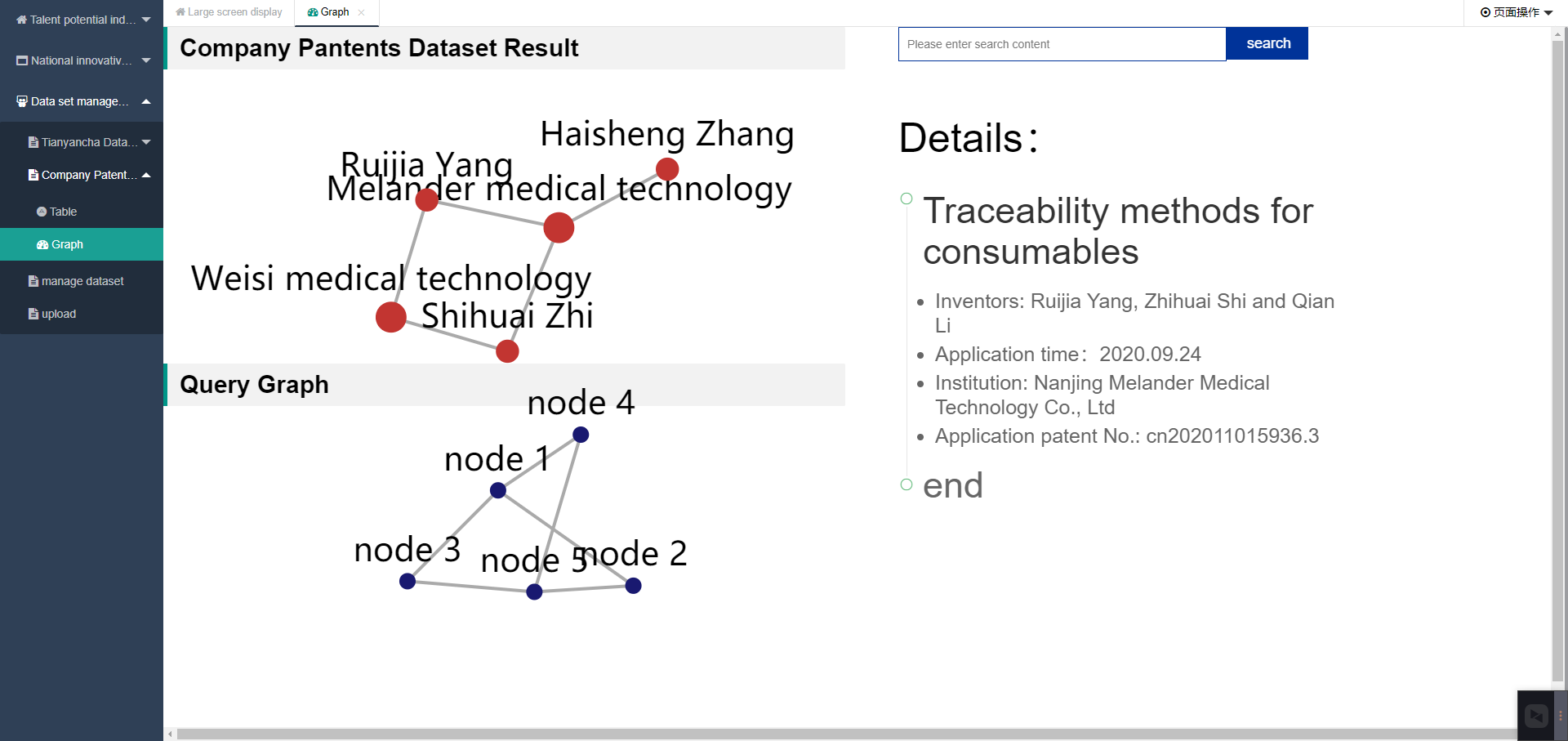}
\vspace{0.2cm}
\caption{Company Patents dataset result and detailed information.}
\end{figure}

\vspace{-1.5mm}
\subsection{Running Time Analysis}
\begin{figure}[h]
\vspace{-3mm}
\centering
\setlength{\abovecaptionskip}{0.5cm}
\includegraphics[width=8cm,height=4.8cm]{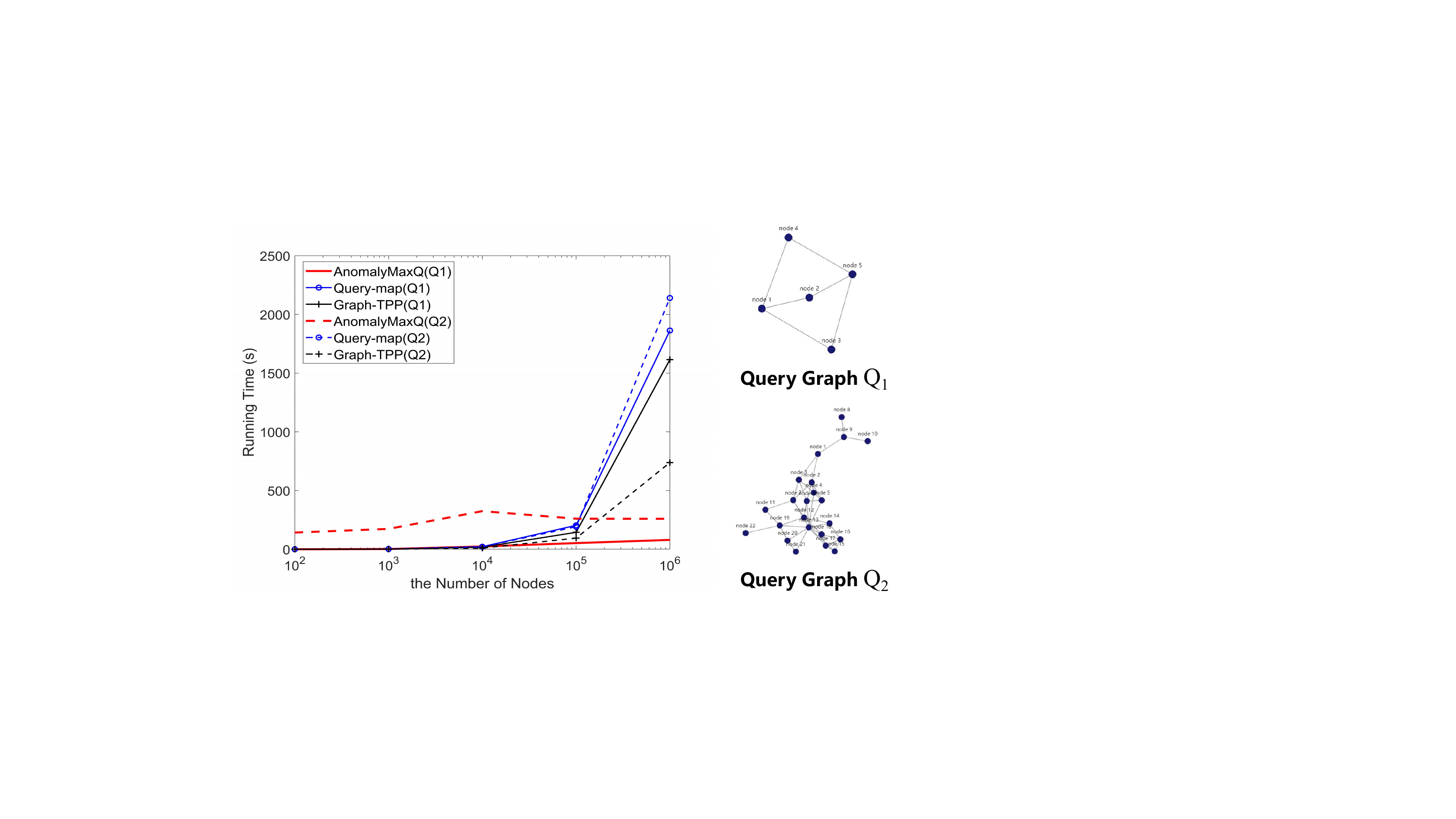}
\vspace{-2mm}
\caption{Comparison of running time with competitive methods.}
\vspace{-5mm}
\end{figure}
% \vspace{3mm}
In this part, we compare the running time of our algorithm with that of ``Graph-TPP'' and ``Query-Map'' \cite{graphtpp,querymap}. Although TSPSD runs very fast, the precision is not satisfactory. The dataset is a simulated ED Dataset. We set up two query graphs, $Q_{1}$ has five nodes, and $Q_{2}$ has 22 points and 92 edges \cite{2012Efficient}.
% Figure 4 reports the comparison between our methods AnomalyMaxQ and the baseline methods on the running time. 
% In Figure 4 the running times were collected from the computer with  Intel(R) Xeon(R) W-2123 ( 8 CPU, 3.60GHz ). 
The results  show that our proposed method always ran faster than all the baseline methods. On the other hand, the number of nodes and edges in the query graph is more influential than attributes graph.
% \vspace{-3mm}
\vspace{-0.5mm}
\section{Conclusion}
In this paper, we introduce \textsc{AnomalyMaxQ} that is capable of predicting business risks and assisting investment decisions. A large number of experiments show that our algorithm has good scalability and fast computing speed, and it only needs 90 seconds response time to run on 7.72m nodes. 
% We are currently improving its functionality and documenting the technical aspects of the system. 
In the future work, we will use multi-dimensional data to generate query graphs automatically.

\section{Acknowledgments}
This work was supported by National Key R$\&$D Program of China (No.2018YFC0832103), partly supported by NSFC (No.61902279), and  China Postdoctoral Science Foundation (No.2019M650048). %The corresponding author is Nannan Wu.

\bibliographystyle{named}
\small
\bibliography{ijcai21}

\begin{thebibliography}{}

\bibitem[\protect\citeauthoryear{Akoglu \bgroup \em et al.\egroup
  }{2014}]{Anomaly_Detection}
Leman Akoglu, Hanghang Tong, and Danai Koutra.
\newblock Graph-based anomaly detection and description: A survey.
\newblock {\em Data Mining and Knowledge Discovery}, 29, 04 2014.

\bibitem[\protect\citeauthoryear{Berk and Jones}{1979}]{BJ}
Robert~H. Berk and Douglas~H. Jones.
\newblock Goodness of fit test statistics that dominate the kolmogorov
  statistics.
\newblock {\em Probability Theory \& Related Fields}, 47(1):47--59, 1979.

\bibitem[\protect\citeauthoryear{Bhattarai \bgroup \em et al.\egroup
  }{2019}]{2019CICE}
Bibek Bhattarai, Hang Liu, and H.~Howie Huang.
\newblock Ceci: Compact embedding cluster index for scalable subgraph matching.
\newblock In {\em Proceedings of the 2019 International Conference on
  Management of Data}, SIGMOD '19, page 1447–1462, New York, NY, USA, 2019.

\bibitem[\protect\citeauthoryear{Donoho and Jin}{2004}]{HC}
David Donoho and Jiashun Jin.
\newblock Higher criticism for detecting sparse heterogeneous mixtures.
\newblock {\em The Annals of Statistics}, 32(3), 2004.

\bibitem[\protect\citeauthoryear{Goodfellow \bgroup \em et al.\egroup
  }{2016}]{deeplearning}
Ian Goodfellow, Yoshua Bengio, and Aaron Courville.
\newblock {\em Deep Learning}.
\newblock MIT Press, 2016.
\newblock \url{http://www.deeplearningbook.org}.

\bibitem[\protect\citeauthoryear{Han \bgroup \em et al.\egroup
  }{2013}]{Turboiso}
Wook-Shin Han, Jinsoo Lee, and Jeong-Hoon Lee.
\newblock Turboiso: towards ultrafast and robust subgraph isomorphism search in
  large graph databases.
\newblock In {\em Proceedings of the 2013 ACM SIGMOD International Conference
  on Management of Data}, pages 337--348, 2013.

\bibitem[\protect\citeauthoryear{Han \bgroup \em et al.\egroup
  }{2019}]{EfficientSubgraphMatching}
Myoungji Han, Hyunjoon Kim, Geonmo Gu, Kunsoo Park, and Wook-Shin Han.
\newblock Efficient subgraph matching: Harmonizing dynamic programming,
  adaptive matching order, and failing set together.
\newblock In {\em Proceedings of the 2019 International Conference on
  Management of Data}, SIGMOD '19, page 1429–1446, New York, NY, USA, 2019.

\bibitem[\protect\citeauthoryear{He and Singh}{2006}]{Closure_Tree}
Huahai He and A.K. Singh.
\newblock Closure-tree: An index structure for graph queries.
\newblock volume~0, pages 38-- 38, 2006.

\bibitem[\protect\citeauthoryear{Jiang \bgroup \em et al.\egroup
  }{2007}]{GString}
Haoliang Jiang, Haixun Wang, Philip Yu, and Shuigeng Zhou.
\newblock Gstring: A novel approach for efficient search in graph databases.
\newblock pages 566--575, 05 2007.

\bibitem[\protect\citeauthoryear{Lai \bgroup \em et al.\egroup
  }{2019}]{2019DistributedSubgraph}
Longbin Lai, Zhu Qing, Zhengyi Yang, Xin Jin, Zhengmin Lai, Ran Wang, Kongzhang
  Hao, Xuemin Lin, Lu~Qin, Wenjie Zhang, Ying Zhang, Zhengping Qian, and
  Jingren Zhou.
\newblock Distributed subgraph matching on timely dataflow.
\newblock {\em Proc. VLDB Endow.}, 12(10):1099–1112, June 2019.

\bibitem[\protect\citeauthoryear{Liu \bgroup \em et al.\egroup
  }{2018}]{2018Futureframe}
Wen Liu, Weixin Luo, Dongze Lian, and Shenghua Gao.
\newblock Future frame prediction for anomaly detection--a new baseline.
\newblock In {\em Proceedings of the IEEE Conference on Computer Vision and
  Pattern Recognition}, pages 6536--6545, 2018.

\bibitem[\protect\citeauthoryear{Liu \bgroup \em et al.\egroup
  }{2019}]{G_Finder}
Lihui Liu, Boxin Du, Jiejun Xu, and Hanghang Tong.
\newblock G-finder: Approximate attributed subgraph matching.
\newblock In {\em 2019 IEEE International Conference on Big Data (Big Data)},
  2019.

\bibitem[\protect\citeauthoryear{Neill}{2012}]{LTSS}
D.~B. Neill.
\newblock Fast subset scan for spatial pattern detection.
\newblock {\em Journal of the Royal Statistical Society, Series B. Statistical
  Methodology}, 2012.

\bibitem[\protect\citeauthoryear{Rivero and
  Jamil}{2017}]{2017efficientAndScalable}
Carlos~R Rivero and Hasan~M Jamil.
\newblock Efficient and scalable labeled subgraph matching using sgmatch.
\newblock {\em Knowledge and Information Systems}, 51(1):61--87, 2017.

\bibitem[\protect\citeauthoryear{{Sun} and {Luo}}{2020}]{2020SubgraphMatching}
S.~{Sun} and Q.~{Luo}.
\newblock Subgraph matching with effective matching order and indexing.
\newblock {\em IEEE Transactions on Knowledge and Data Engineering}, pages
  1--1, 2020.

\bibitem[\protect\citeauthoryear{Sun \bgroup \em et al.\egroup
  }{2012a}]{2012Efficient}
Z.~Sun, H.~Wang, H.~Wang, B.~Shao, and J.~Li.
\newblock Efficient subgraph matching on billion node graphs.
\newblock {\em Proceedings of the Vldb Endowment}, 5(9):788--799, 2012.

\bibitem[\protect\citeauthoryear{Sun \bgroup \em et al.\egroup
  }{2012b}]{Efficient_subgraph_matching}
Zhao Sun, Hongzhi Wang, Haixun Wang, Bin Shao, and Jianzhong Li.
\newblock Efficient subgraph matching on billion node graphs.
\newblock {\em arXiv preprint arXiv:1205.6691}, 2012.

\bibitem[\protect\citeauthoryear{Sun \bgroup \em et al.\egroup
  }{2020}]{DBLP:conf/cikm/00050WYC20}
Ying Sun, Wenjun Wang, Nannan Wu, Wei Yu, and Xue Chen.
\newblock Anomaly subgraph detection with feature transfer.
\newblock In Mathieu d'Aquin, Stefan Dietze, Claudia Hauff, Edward Curry, and
  Philippe Cudr{\'{e}}{-}Mauroux, editors, {\em {CIKM} '20: The 29th {ACM}
  International Conference on Information and Knowledge Management, Virtual
  Event, Ireland, October 19-23, 2020}, pages 1415--1424. {ACM}, 2020.

\bibitem[\protect\citeauthoryear{Taha and Yoo}{2015}]{SIIMCO}
Kamal Taha and Paul~D Yoo.
\newblock Siimco: A forensic investigation tool for identifying the influential
  members of a criminal organization.
\newblock {\em IEEE Transactions on Information Forensics and Security},
  11(4):811--822, 2015.

\bibitem[\protect\citeauthoryear{Willett \bgroup \em et al.\egroup
  }{1998}]{Willett1998Chemical}
Peter Willett, John~M. Barnard, and Geoffrey~M. Downs.
\newblock Chemical similarity searching. jchem inf comput sci.
\newblock {\em Journal of Chemical Information \& Modeling}, 38(6):983--996,
  1998.

\bibitem[\protect\citeauthoryear{Wu \bgroup \em et al.\egroup }{2016}]{TSPSD}
Nannan Wu, Feng Chen, Jianxin Li, Baojian Zhou, and Naren Ramakrishnan.
\newblock Efficient nonparametric subgraph detection using tree shaped priors.
\newblock In {\em Proceedings of the AAAI Conference on Artificial
  Intelligence}, volume~30, 2016.

\bibitem[\protect\citeauthoryear{Wu \bgroup \em et al.\egroup
  }{2017}]{graphtpp}
Nannan Wu, Feng Chen, Jianxin Li, Jinpeng Huai, and Bo~Li.
\newblock Query-driven discovery of anomalous subgraphs in attributed graphs.
\newblock pages 3105--3111, 08 2017.

\bibitem[\protect\citeauthoryear{Wu \bgroup \em et al.\egroup
  }{2019a}]{DBLP:journals/tkde/WuCLHZLR19}
Nannan Wu, Feng Chen, Jianxin Li, Jinpeng Huai, Baojian Zhou, Bo~Li, and Naren
  Ramakrishnan.
\newblock A nonparametric approach to uncovering connected anomalies by tree
  shaped priors.
\newblock {\em {IEEE} Trans. Knowl. Data Eng.}, 31(10):1849--1862, 2019.

\bibitem[\protect\citeauthoryear{Wu \bgroup \em et al.\egroup
  }{2019b}]{querymap}
Nannan Wu, Wenjun Wang, Feng Chen, Jianxin Li, Bo~Li, and Jinpeng Huai.
\newblock Uncovering specific-shape graph anomalies in attributed graphs.
\newblock {\em Proceedings of the AAAI Conference on Artificial Intelligence},
  33:5433--5440, 07 2019.

\bibitem[\protect\citeauthoryear{Yan \bgroup \em et al.\egroup
  }{2004}]{Graph_Indexing}
Xifeng Yan, Philip Yu, and Jiawei Han.
\newblock Graph indexing: A frequent structure-based approach.
\newblock {\em Proceedings of the ACM SIGMOD International Conference on
  Management of Data}, 06 2004.

\bibitem[\protect\citeauthoryear{Yang and Sze}{2007}]{Yang2007Path}
Qingwu Yang and Sing~Hoi Sze.
\newblock Path matching and graph matching in biological networks.
\newblock {\em Journal of Computational Biology}, 14(1):56--67, 2007.

\bibitem[\protect\citeauthoryear{Zhao \bgroup \em et al.\egroup
  }{2020a}]{DBLP:journals/tkde/ZhaoCY20}
Liang Zhao, Feng Chen, and Yanfang Ye.
\newblock Efficient learning with exponentially-many conjunctive precursors for
  interpretable spatial event forecasting.
\newblock {\em {IEEE} Trans. Knowl. Data Eng.}, 32(10):1923--1935, 2020.

\bibitem[\protect\citeauthoryear{Zhao \bgroup \em et al.\egroup
  }{2020b}]{DBLP:journals/geoinformatica/ZhaoCCJWLR20}
Liang Zhao, Jiangzhuo Chen, Feng Chen, Fang Jin, Wei Wang, Chang{-}Tien Lu, and
  Naren Ramakrishnan.
\newblock Online flu epidemiological deep modeling on disease contact network.
\newblock {\em GeoInformatica}, 24(2):443--475, 2020.

\bibitem[\protect\citeauthoryear{Zou \bgroup \em et al.\egroup
  }{2007}]{Top-k_subgraph_matching}
Lei Zou, Lei Chen, and Yansheng Lu.
\newblock Top-k subgraph matching query in a large graph.
\newblock pages 139--146, 01 2007.

\end{thebibliography}

\end{document}